\documentclass[prd,twocolumn,showpacs,reprint,preprintnumbers,nofootinbib,superscriptaddress]{revtex4-1}

\input{header}

\newcommand{\orcid}[1]{\begingroup
  \hypersetup{hidelinks}\href{https://orcid.org/#1}{\includegraphics[width=10pt]{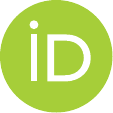}} \endgroup}
  
\begin{document}

\title{{\Large Neutrino Tridents and $W$-boson production at the Muon Collider}}

\author{F. Kling\,\orcidlink{0000-0002-3100-6144}}%
\email{fkling@uni-bonn.de}%
\affiliation{Universit\"at Bonn, Regina-Pacis-Weg 3, D-53113 Bonn, Germany}%

\author{T. M\"akel\"a\orcidlink{0000-0002-1723-4028}}%
\email{tmakela@uci.edu}%
\affiliation{Department of Physics and Astronomy, University of California, 92697, Irvine, CA, USA}%

\author{Sebastian Trojanowski\orcidlink{0000-0003-2677-0364}}
\email{strojanowski@camk.edu.pl}
\affiliation{National Centre for Nuclear Research, Pasteura 7, Warsaw, 02-093, Poland}
\affiliation{Astrocent, Nicolaus Copernicus Astronomical Center Polish Academy of Sciences, ul. Rektorska 4, 00-614, Warsaw, Poland}

\author{Bei Zhou \orcid{0000-0003-1600-8835}\,}
\email{beizhou@psu.edu}
\affiliation{Department of Physics, The Pennsylvania State University, University Park, Pennsylvania 16802, USA}
\affiliation{Theory Division, Fermi National Accelerator Laboratory, Batavia, Illinois 60510, USA}
\affiliation{Kavli Institute for Cosmological Physics, University of Chicago, Chicago, Illinois 60637, USA} 


\begin{abstract}

Muon colliders are exceptionally intense sources of multi-TeV neutrinos, allowing even relatively small detectors to record large rates of neutrino interactions. In addition to serving as neutrino-ion collider for precision measurements of neutrino deep inelastic scattering, this setup allow to study even rare neutrino processes with at high statistics. In this study, we investigate the feasibility of measuring neutrino trident and on-shell W-boson production, predicting over one million trident events and one hundred thousand W-boson events in a 100 kg silicon-based active target over ten years and allowing to detect these processes for the first time. We further show that these measurements could constrain new vector and axial-vector interactions beyond projected DUNE sensitivities and provide unprecedented constraints on the high-x photon distribution through W-boson production near threshold. We also discuss strategies for signal-background discrimination and the resulting detector requirements.
\end{abstract}

\maketitle 
\clearpage

\section{Introduction} 
\label{sec:introduction}

The Standard Model (SM) of particle physics has been extensively tested and firmly established, yet the unresolved properties of neutrinos continue to provide an exceptional window into new physics. Among the rare and most sensitive probes of the electroweak sector is neutrino trident production, a process in which a neutrino scatters in the Coulomb field of a heavy nucleus to produce a pair of charged leptons, $\nu N \to \nu^{(\prime)} \ell^- \ell^{(\prime)+} N$~\cite{Kozhushner1961, Shabalin1963, Czyz:1964zz, Lovseth:1971vv, Fujikawa:1971nx, Koike:1971tu, Koike:1971vg, Brown:1972vne, Brown:1973ih, Belusevic:1987cw, Altmannshofer:2014pba, Magill:2016hgc, Ge:2017poy, Ballett:2018uuc, Altmannshofer:2019zhy, Gauld:2019pgt, Zhou:2019vxt, Zhou:2019frk}. The dominant production mode is coherent scattering, in which the target nucleus remains intact. This leads to a cross section enhanced by the square of the nuclear charge ($Z^2$) and yields a distinct event topology without additional nuclear-recoil products. 

Although several lepton combinations are possible, the dimuon channel, $\nu_\mu N \to \nu_\mu \mu^+ \mu^- N$, is the most experimentally accessible. This channel provides a distinct signature, as both outgoing muons can be identified and measured. In the SM, dimuon pair production from an incident muon neutrino proceeds via both charged-current ($W$-boson) and neutral-current ($Z$-boson) exchanges. Destructive interference between the two lowers the total SM cross section, making the process particularly sensitive to the electroweak gauge couplings.
  
Consequently, neutrino tridents are recognized as a promising laboratory for discovering physics beyond the Standard Model (BSM). For instance, a new light vector mediator, such as a $Z^\prime$ gauge boson associated with an $L_\mu - L_\tau$ gauge symmetry~\cite{Altmannshofer:2014cfa,Altmannshofer:2014pba}, or new heavy degrees of freedom parameterized by effective operators~\cite{Altmannshofer:2019zhy}, can bypass this destructive interference. Such new physics would significantly increase the trident production rate and alter the kinematic distributions of the outgoing leptons. 

Conclusive, unambiguous observation of neutrino tridents remains experimentally challenging. This stems from the strongly suppressed cross section, typically several orders of magnitude smaller than standard charged-current (CC) deep inelastic scattering (DIS), and from the difficulty of mitigating backgrounds. Historically, the CHARM-II~\cite{CHARM-II:1990dvf}, CCFR~\cite{CCFR:1991lpl}, and NuTeV~\cite{NuTeV:1999wlw} collaborations conducted searches for dimuon neutrino tridents using neutrino beams with energies up to $160~\text{GeV}$. While earlier measurements reported evidence at about the $3\sigma$ level, subsequent studies highlighted severe background challenges, most notably from diffractive charm production. This lack of a definitive measurement has motivated renewed interest in searching for tridents at modern and future experimental facilities, including the Deep Underground Neutrino Experiment (DUNE)~\cite{Altmannshofer:2019zhy,Bigaran:2025vea}, forward neutrino detectors at the Large Hadron Collider (LHC)~\cite{Altmannshofer:2024hqd,Francener:2024wul}, and neutrino telescopes and other accelerator facilities~\cite{Altmannshofer:2014pba,Magill:2016hgc,Ge:2017poy,Ballett:2018uuc,Altmannshofer:2019zhy,Zhou:2019vxt,Zhou:2019frk,Sarkar:2023dvr,Choi:2026yzw}.

In this paper, we present the first comprehensive study of rare, photon-mediated neutrino interactions at a future multi-TeV muon collider, including both non-resonant neutrino trident production and (on-shell) $W$-boson production (WBP), $\nu_\ell N \to \ell^- W^+ N^{(\prime)}$. Leveraging the exceptionally high flux of highly energetic neutrinos from muon decays, a dedicated forward silicon neutrino detector could enable high-significance measurements of these rare electroweak processes for the first time; see also Refs~\cite{Batell:2024cdl,Das:2026eyy} for a discussion of muon-induced tridents at the muon collider. Recent estimates of trident production at the muon collider have focused on the coherent and diffractive regimes~\cite{Choi:2026yzw,Francener:2026ukt}. However, the multi-TeV neutrino energies available at such a facility will also exceed the kinematic threshold required to produce an on-shell intermediate $W$ boson in the charged-current trident diagram. Hence, this experimental setup offers a unique opportunity to measure the WBP process, which has never been possible, and to indirectly constrain fundamental hadronic structure uncertainties, such as the large-$x$ photon parton distribution function (PDF) of the nucleon~\cite{Xie:2021equ,Xie:2023qbn}.

The WBP followed by a leptonic $W$ decay may provide a significant resonant enhancement of the inclusive trident process \cite{Zhou:2019vxt}. This resonant regime leads to new event topologies: leptonic $W$-boson decays yield ``resonant tridents'' characterized by a large dimuon opening angle and soft hadronic recoil, while hadronic $W$ decays produce fat-jets with a distinctive substructure. We discuss how a dedicated forward neutrino detector at a muon collider could successfully isolate these events. 

\begin{figure*}
    \centering
    \includegraphics[width=0.99\linewidth]{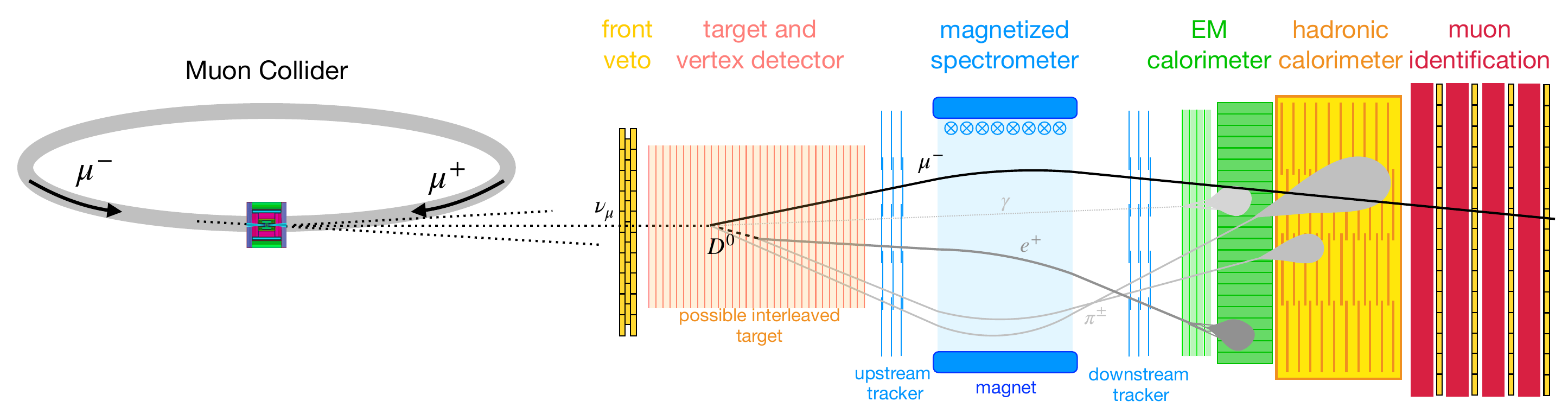}
    \caption{Schematic of a forward neutrino detector, MuCol$\nu$, placed inside the muon collider neutrino beam.}
    \label{fig:mucolv}
\end{figure*}

The paper is structured as follows. \Cref{sec:nuatmuc} outlines the properties of forward neutrinos at the muon collider and briefly presents the detector concept. \Cref{sec:numutridents} discusses the non-resonant neutrino trident production in this context. In \cref{sec:Wboson}, we describe the closely related process of WBP and its direct connection to the trident signal. \Cref{sec:BSMbounds} explores the bounds on BSM physics derived from these measurements, and we conclude in \cref{sec:conclusions}.

\section{Neutrinos at the Muon Collider} 
\label{sec:nuatmuc}

The muon collider consists of a complex sequence of accelerator elements to produce, cool, and accelerate the muon beam. At the end of this chain is a circular storage ring, where the beam is brought into collision at the main interaction points, and the resulting particles are studied. Since muons have a finite lifetime of 2.2~$\mu$s, the stored muons decay continuously while circulating the ring. This makes a muon collider an exceptionally bright source of energetic neutrinos. 

In this study, we consider a 10~TeV center-of-mass energy muon collider, as considered for the European Strategy Update~\cite{InternationalMuonCollider:2025sys}. This design includes a 10~km-long storage ring, into which approximately $10^{13}$ muons are injected every second~\cite{MuCoL:2024oxj, MuCoL:2025quu}.  Although muon decays occur throughout the storage ring, the largest neutrino flux is generated by decays in the straight sections located around the interaction points. In the considered design, the straight section of the tunnel has a length of about 500~m. About $5\times 10^{11}$ muon neutrinos and electron anti-neutrinos per second are produced from the decay of the $\mu^-$ beam, forming an intense and strongly collimated beam of high-energy neutrinos. A detailed simulation of the neutrino flux, taking into account the trajectories of muons through the muon collider's interaction region, has been obtained in Refs.~\cite{Burk:2026fox, Choi:2026yzw}. In the following, we will use the flux of Ref.~\cite{Burk:2026fox}.

Unlike the ongoing LHC~\cite{Feng:2022inv, Ariga:2025qup, Zhou:2025iso} and proposed FCC~\cite{MammenAbraham:2024gun} neutrino program, the properties of the forward neutrino beam at the muon collider are well understood. Refs.~\cite{Burk:2026fox, Choi:2026yzw} showed that contributions from other sources, including muon collisions at the primary interaction point, electromagnetic showers caused by Michel electrons, and secondary neutrino interactions, are suppressed by about a factor of $10^{10}$ compared to the primary flux, making the muon collider's neutrino beam remarkably pure. The energy and angular spectra of neutrinos from muon decays can be simulated precisely.

Studying neutrino interactions offers the opportunities to perform a variety of measurements and searches with unprecedented sensitivity, for example probing proton structure~\cite{InternationalMuonCollider:2025sys}, performing electroweak precision measurements~\cite{Marzocca:2025inb, deGouvea:2025zfq}, or searching for signs of new physics~\cite{Adhikary:2024tvl, Kling:2025zsb}. To exploit this opportunity, we consider the MuCol$\nu$ detector design outlined in Ref.~\cite{Burk:2026fox}. The detector is located about 500~m downstream of the main experiment. The distance was selected to optimize the expected event yield while keeping the detector at least 10~m away from the muon beam, providing the necessary room to accommodate the experimental setup and shield it from beam-induced backgrounds. 

Because of the strong collimation and high beam intensity, even a small detector at this position will record a large event rate. Following Ref.~\cite{Burk:2026fox}, we consider a cylindrical target of radius 25~cm and mass of 100~kg, using silicon as the target material. Fig.~\ref{fig:mucolv} shows a proposed design for the MuCol$\nu$ detector. Located at the upstream part of the detector is a front veto station followed by a target made of several high-precision tracking layers, which also serve as a vertex detector, as originally proposed in Ref.~\cite{King:1997dx}. The excellent spatial resolution of these layers allows the primary interaction point to be reconstructed and enables the identification of displaced secondary vertices. This provides sensitivity to the decays of relatively short-lived particles, including charm hadrons and tau leptons. The neutrino interaction rate could be increased substantially by adding layers of dense material between the active tracking elements, thereby increasing the target mass. We therefore additionally present event rate estimates for a tungsten target.

The downstream section contains a magnetized spectrometer based on the FASER2 conceptual design~\cite{Adhikary:2024nlv, Salin:2927003, FPF:2025bor}. It consists of a dipole magnet surrounded by tracking stations placed before and after the magnetic field, allowing high-precision measurement of the charge and momentum of multi-TeV particles. The spectrometer is followed by electromagnetic and hadronic calorimeters, with a dedicated muon identification system completing the detector.

At the energies of interest, neutrinos interact primarily through charged-current deep inelastic scattering; we estimate an event rate of $1.1 \cdot 10^9$ muon neutrino and $5.2 \cdot 10^8$ anti-electron neutrino interactions per year in this target. In addition, the large neutrino flux provides an opportunity to probe even rare interaction processes with high statistics~\cite{Choi:2026yzw}.

\section{Non-Resonant Neutrino Tridents}\label{sec:numutridents}

\begin{figure*}[t]
    \centering
    \includegraphics[width=0.24\textwidth]{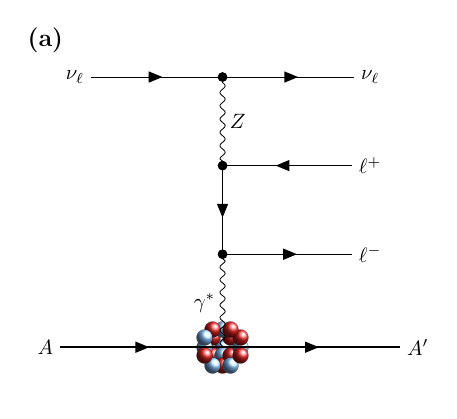}\hfill
    \includegraphics[width=0.24\textwidth]{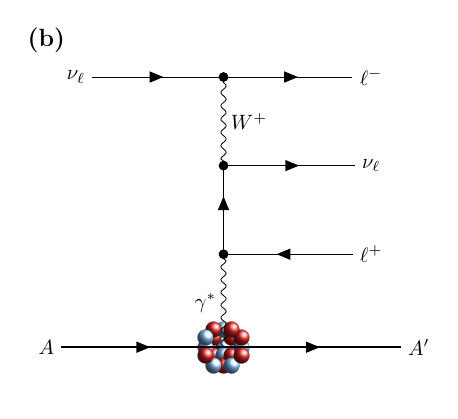}\hfill
    \includegraphics[width=0.24\textwidth]{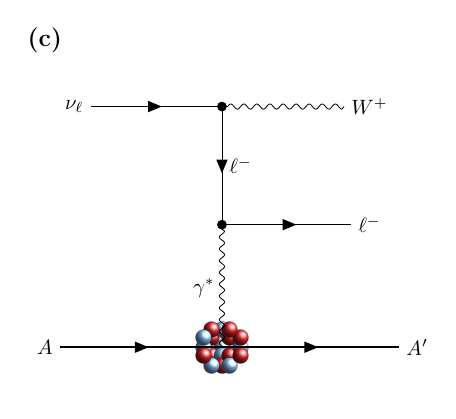}\hfill
    \includegraphics[width=0.24\textwidth]{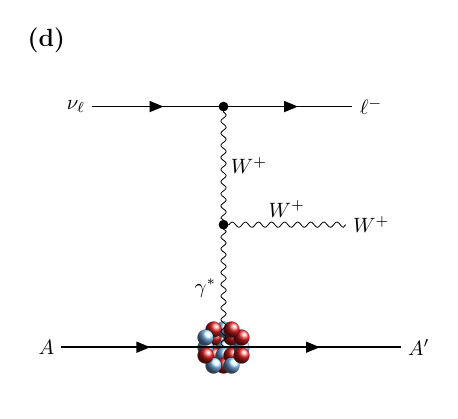}
\caption{Illustrative Feynman diagrams for rare neutrino interactions in the Coulomb field of a nucleus: (a) non-resonant neutrino trident production via neutral-current $Z$ exchange (which can also be mediated by new physics like a $Z^\prime$); (b) non-resonant trident production via charged-current $W$ exchange; (c) WBP where the virtual photon couples to the final-state charged lepton; (d) WBP employing the triple gauge coupling where the virtual photon couples directly to the $W$ boson. In the inclusive picture, leptonic $W$ decays in (c) and (d) provide the resonant enhancement of the charged-current trident channel.}
    \label{fig:diagrams}
\end{figure*}

The highly collimated flux of TeV-scale neutrinos generated in the straight sections of the muon collider offers an extraordinary opportunity to study rare electroweak interactions, particularly neutrino trident production. We first focus on the non-resonant regime of this process, see \cref{fig:diagrams}\textcolor{blue}{(a)}\textcolor{black}\ and \textcolor{blue}{(b)}\textcolor{black}, which facilitates purely coherent trident production where the target nucleus remains intact. Because the coherent trident cross section grows linearly with incident neutrino energy before saturating at the electroweak scale~\cite{Zhou:2019vxt}, TeV-scale neutrinos yield substantially higher interaction rates than past fixed-target experiments~\cite{Choi:2026yzw}. In addition, the initial state of the muon collider neutrino beam is characterized by a well-known flavor composition (e.g., dominated by $\nu_\mu$ and $\bar{\nu}_e$ from $\mu^-$ decays). This avoids the complex hadronic flux uncertainties typical for conventional proton-driven beams, which affect the LHC's potential to measure trident cross sections~\cite{Altmannshofer:2024hqd}.

The expected non-resonant trident event rates are summarized in \cref{tab:event_rates} for the baseline $100~\text{kg}$ silicon and tungsten targets over a 10-year run. The dimuon channel, $\nu_\mu N\to \nu_\mu \mu^+\mu^-N$, is the most experimentally accessible and is expected to produce around $\textrm{a few} \times 10^5$ events. Other, more challenging final states can also be studied, with similar projections for the $\mu^-\tau^+$ channel and about $2 \times 10^4$ events for $\tau^-\tau^+$. While event counts with a pure silicon detector are already impressive, adding passive, higher-$Z$ target materials interleaved with active tracking elements would further increase the expected scattering rate. This would make the future muon collider particularly well-suited for studying neutrino trident processes.

\begin{table*}[htbp]
\centering
\caption{Expected number of coherent non-resonant neutrino trident interactions per visible final state for benchmark $100~\text{kg}$ silicon (radius $25~\text{cm}$, length $21.9~\text{cm}$) and tungsten (length $2.6~\text{cm}$) targets in ten years of data taking. The total number of inclusive charged-current deep inelastic scattering (CC DIS) events is shown for comparison. For visible finals states where both $\nu_\mu$ and $\overline{\nu}_e$ contribute, the incoming neutrino flavor is summed over.}
\label{tab:event_rates}
\resizebox{\textwidth}{!}{%
\begin{tabular}{c|c|c||c|c|c|c|c|c|c}
\hline
\hline
Target & Mass & Length & \multicolumn{7}{c}{Event rates per visible final state}\\ 
 nucleus & [kg] & [cm] & $e^- e^+$ & $e^+ \mu^-$ & $e^+ \tau^-$ & $\mu^- \mu^+$ & $\mu^- \tau^+$ & $\tau^- \tau^+$ & CC DIS\\ 
\hline
Si & 100 & $21.9$ & $4.67\times 10^5$ & $1.10\times 10^6$ & $1.55\times 10^5$ & $3.05\times 10^5$ & $1.41\times 10^5$ & $1.57\times 10^4$ & $1.98\times 10^{10}$\\ 
\hline
W & 100 & 2.6 & $1.25\times 10^6$ & $2.80\times 10^6$ & $3.34\times 10^5$ & $7.02\times 10^5$ & $2.73\times 10^5$ & $2.57\times 10^4$ & $2.04\times 10^{10}$\\ 
\hline
\hline
\end{tabular}%
}
\end{table*}

Operating in this high-statistics environment, isolating the trident signal will rely on rejecting the overwhelming DIS backgrounds. The significant boost of the TeV-scale neutrinos gives neutrino tridents a distinct topological signature: a pair of high-energy, forward-going leptons emerging from a single isolated vertex with minimal hadronic activity. Taking advantage of the exceptional spatial, angular, and energy resolution of the considered forward neutrino detector, one can require the exclusive detection of exactly two signal leptons with no missing particles, besides the outgoing neutrinos. For the dimuon channel, requiring exactly two muons effectively removes most ordinary DIS interactions.

However, certain rare event topologies can mimic the pure leptonic trident signature and must be suppressed. The main sources of background are interactions that produce a primary muon and a charged pion, $\nu_\ell N \to \mu^- \pi^+ X$, where the pion may decay in flight, $\pi^+ \to \mu^+ \nu_\mu$, before reaching the calorimeter or muon detection system. At multi-TeV energies, these interactions are dominated by CC DIS, where the hadronic remnant $X$ consists primarily of neutral particles~\cite{Altmannshofer:2024hqd}. These backgrounds, however, may be actively vetoed in the MuCol$\nu$ detector by identifying secondary interactions of these neutral hadrons. This leaves smaller, more exclusive contributions from coherent and resonant pion production, which lack significant hadronic activity and feature either a distinctly forward-scattered pion with no nuclear recoil or a single pion with a recoiling nucleon.

In resonant charged-current single-pion production (CCRES), $\nu_\mu N \to \mu^- N^* \to \mu^- N' \pi^+$, the hadronic invariant mass is restricted to the resonance region, $W \lesssim 2~\text{GeV}$. As a result, the outgoing muon carries away most of the incident TeV-scale neutrino energy, resulting in a relatively soft pion, $E_\pi \lesssim \mathcal{O}(10)~\text{GeV}$. In addition, a recoiling nucleon may produce additional hadronic activity at the interaction vertex, depending on final-state interaction (FSI) effects. These signal characteristics allow for efficient active vetoing in the silicon target.

Instead, coherent CC pion production, $\nu_\mu A \to \mu^- \pi^+ A$, leaves the target nucleus in its ground state without a measurable nuclear recoil in the silicon target. At high energies, the coherent cross section remains approximately energy independent, $\sigma_{\text{COH}} \sim \text{a few} \times 10^{-39}~\text{cm}^2$ per nucleus~\cite{Formaggio:2012cpf}. For our benchmark $100~\text{kg}$ silicon target (radius $R = 25~\text{cm}$, length $L = 21.9~\text{cm}$), this cross section corresponds to predicted $\sim 2 \times 10^5$ coherent pion interactions over the full 10-year exposure.

To determine whether these coherent pions can mimic a clean dimuon topology, we consider both pion rescattering within the active silicon target and decay in flight in the downstream detector volume. Assuming that the pion can be produced uniformly in any part of the active target volume, the probability that it exits the active silicon volume without interacting hadronically and without decaying is given by $P_{\text{escape}} = (\lambda_{\text{eff}}/L) \left( 1 - e^{-L / \lambda_{\text{eff}}} \right)$, where $\lambda_{\text{eff}}^{-1} = \lambda_{\text{int}}^{-1} + \lambda_{\text{dec}}^{-1}$. The pion interaction length in silicon is $\lambda_{\text{int}} \approx 59.1~\text{cm}$, while the pion decay length is $\lambda_{\text{dec}} = \gamma c \tau_\pi \approx 5.6~\text{km} \times (E_\pi / 100~\text{GeV})\gg \lambda_{\text{int}}$~\cite{ParticleDataGroup:2026mpi}, yielding $P_{\text{escape}} \approx 84\%$ for a $100~\textrm{GeV}$ pion. For the pions escaping the active target, we estimate the probability of decaying in flight over a downstream tracking distance of $L_{\text{track}} \sim 10~\text{m}$, $P_{\text{dec}} = L_{\text{track}} / \lambda_{\text{dec}} \sim 10^{-3}$, for $E_{\pi}\sim 100~\text{GeV}$. Consequently, the initial coherent pion rate is suppressed to $\sim 10^2-10^3$ potential background events over the 10-year run. These remaining backgrounds can be further suppressed by cuts on the opening angle between the two charged tracks, identifying a characteristic tracking kink caused by the $\pi \to \mu$ decay in flight, as well as by considering the energy loss and rescattering probability of charged pions in the downstream detector material~\cite{Altmannshofer:2024hqd}.

Another prompt background comes from charmed meson production, such as $\nu_\mu N \to \mu^- D^+ X$, followed by the semi-leptonic decay $D^+ \to \mu^+ \nu_\mu X^\prime$~\cite{Faura:2020oom, Zhou:2021xuh}. As with misreconstructed pions, a charged-track multiplicity cut, i.e., exactly two emerging tracks, effectively rejects most inelastic DIS events and neutral $D^0$ meson decays in charmed-induced backgrounds, which are typically accompanied by additional charged hadronic products. In the active-target detector MuCol$\nu$, rescattering of neutral hadrons can also veto remaining DIS events with no charged hadrons or $D^+$ meson decay. These active tracking and timing capabilities give MuCol$\nu$ an advantage over passive emulsion detectors like FASER$\nu$, which must rely on, e.g., complex reverse-tracking methods to isolate the trident signal~\cite{Altmannshofer:2024hqd}.

Beyond inclusive DIS processes, a distinct dimuon background results from diffractive charmed strange meson production, $\nu_\mu A \to \mu^- D_s^{(*)+} A$, with the cross section measured by the NuTeV collaboration to be $\sigma_{\text{diff}} = (3.0 \pm 0.8) \times 10^{-39}~\text{cm}^2$ per nucleon at $E_\nu = 130~\text{GeV}$~\cite{NuTeV:1999wlw}. The final-state $D_s^+$ can decay producing a muon primarily via two leptonic modes: the direct two-body decay $D_s^+ \to \mu^+ \nu_\mu$ ($\text{BR} \approx 0.54\%$) and the tauonic cascade $D_s^+ \to \tau^+ \nu_\tau \to \mu^+ \nu_\mu \bar{\nu}_\tau \nu_\tau$ (combined $\text{BR} \approx 0.93\%$)~\cite{ParticleDataGroup:2026mpi}, giving an effective leptonic branching fraction of about $1.5\%$. Notably, the cascade event features an intermediate $\tau$ lepton and produces three missing neutrinos in the final state. This softens the secondary muon momentum, increases missing transverse momentum $p_T^{\text{miss}}$, and produces an additional secondary vertex kink from the $\tau$ decay, leading to distinct event characteristics. The direct decay $D_s^+ \to \mu^+ \nu_\mu$ is thus the most trident-like channel, contributing $\sim 3 \times 10^4$ events over 10 years before selection cuts.

Crucially, diffractive $D_s$ events can be suppressed by combining vertex displacement resolution with kinematic handles, as established by the NuTeV analysis~\cite{NuTeV:1999wlw}:
\begin{enumerate}
    \item \textbf{Displaced Secondary Vertices:} In diffractive scattering, the nucleon recoils elastically with minimal energy transfer, such that $E_{D_s^+} \approx E_\nu - E_\mu$. For neutrino energies at the TeV scale, this results in $D_s^+$ energies of several hundred GeV and Lorentz boosts of $\gamma_{D_s^\pm} \sim 100$. The corresponding laboratory flight distance is extended to macroscopic lengths, $L_{\text{lab}} = \gamma c\tau \sim \text{cm}$. Notably, required vertexing and tracking requirements can be readily achieved with modern high-granularity silicon pixel detectors, which routinely deliver spatial resolutions down to a few micrometers, as demonstrated by the current and planned tracking detector upgrades at the LHC~\cite{Garcia-Sciveres:2017ymt}, cf. also review of tracking technologies to be employed in the proposed muon collider and Future Circular Collider (FCC) facilities~\cite{FCC:2018vvp,FCC:2018evy,InternationalMuonCollider:2025sys}.
    \item \textbf{Invariant Mass and Azimuthal Topology:} Kinematic distributions provide further discrimination of $D_s^+$-induced events. In diffractive production, the $D_s^+$ recoils against the primary muon, forcing the dimuon pair into a back-to-back transverse topology, $\Delta\phi_{\mu\mu} \to \pi$. In contrast, coherent trident muons are produced more symmetrically. Furthermore, the dimuon invariant mass peak for diffractive charm is found at higher masses than the characteristic low-mass threshold peak of coherent tridents, $M_{\mu\mu} \lesssim 1.5~\text{GeV}$.
    \item \textbf{Excited State Decays:} The diffractive process can also produce the excited vector state $D_s^{*+}$, which decays with a $93.6\%$ branching ratio via $D_s^{*+} \to D_s^+ \gamma$~\cite{ParticleDataGroup:2026mpi}. At multi-TeV energies, the large Lorentz boost of the parent $D_s^{*+}$ is expected to shift the energy of this transition photon, $E_\gamma\approx 144~\text{MeV}$ in the meson's rest frame, to several GeV in the laboratory frame. This creates an electromagnetic shower that can veto such events.      
\end{enumerate}

Beyond the $\mu\mu$ tridents, probing trident interactions that contain tau leptons in the final state, such as $\mu^\pm\tau^\mp$ and $\tau^+\tau^-$, would require $\tau$ identification. Importantly, high-energy $\tau$ leptons have an average decay length in the cm range, e.g., $\gamma c\tau\sim 5~\text{cm}$ for $E_\tau\simeq \text{TeV}$. These characteristic displacements make it feasible to identify the secondary $\tau$ decay vertices, provided that the tau decay kink can be measured with sufficient angular resolution, $\theta \sim m_\tau / E_\tau \sim \text{GeV/TeV} \sim 10^{-3}$. Furthermore, while the undetected neutrinos prevent full kinematic reconstruction, the resulting missing transverse momentum provides an additional handle for tau identification. If the tau decays hadronically, the event can easily be confused with ordinary CC DIS or single-pion production backgrounds. Disentangling true $\mu\tau$ or $\tau\tau$ tridents from the much larger backgrounds of CC charm production, $\nu_\mu N \to \mu^- c X$ with $c \to \tau^+\nu_\tau X$, could be achieved by additional kinematic discriminators, such as the ratio of missing transverse momentum to the total visible transverse momentum, $R^T_{\text{Miss}} = p^T_{\text{Miss}} / (p^T_{\text{Miss}} + p^T_\mu)$~\cite{Bigaran:2025vea}. 

Measuring these elusive neutrino trident production channels will thus require high-precision vertexing, excellent missing transverse energy resolution, and advanced event analysis. In terms of detector design, this requires a careful optimization of the spectrometer length: it must be kept at a minimum to suppress the dominant background from in-flight $\pi \to \mu$ decays, yet remain long enough to ensure robust charge identification and precise momentum resolution for tracks up to $5~\text{TeV}$. We note that advanced deep-learning algorithms developed for the LHC, such as those used by the CMS collaboration, currently achieve $\tau$-tagging efficiencies of $\epsilon_\tau \sim 80\%$ while keeping mis-identification rates for light-quark jets, muons, and electrons at the $\mathcal{O}(10^{-3})$ level~\cite{CMS:2022prd}. Implementing such techniques in the cleaner, pileup-free environment of a forward neutrino detector is expected to yield even higher efficiencies, enabling successful $\mu\tau$ and $\tau\tau$ trident measurements in MuCol$\nu$.

\section{$\mathbf{W}$ boson production}
\label{sec:Wboson}

When neutrino energies reach the multi-TeV range, the inclusive trident process receives a substantial resonant enhancement due to the kinematically allowed on-shell production of a $W$ boson, $\nu_\ell + A \to \ell^- + W^+ + X$. The interaction proceeds through the exchange of a virtual photon between the target nucleus and the charged final-state particles ($\ell^-$ or $W^+$), cf. \cref{fig:diagrams}\textcolor{blue}{(c)}\textcolor{black}\ and \textcolor{blue}{(d)}\textcolor{black}. While standard CC DIS relies on weak boson exchange with quarks, WBP depends on the electromagnetic coupling to the nucleus, placing it within the same family of interactions as neutrino tridents~\cite{Zhou:2019vxt,Zhou:2019frk}. Moreover, since the matrix element is invariant under a CP transformation, both neutrinos and antineutrinos have the same total and differential cross sections for this process.

In addition, WBP and non-resonant neutrino trident production are connected at the amplitude level. Above the $W$-production threshold, the CC and CC+NC trident channels are dominated by resonant contribution from on-shell $W$ production followed by the leptonic decay $W^+\to \ell^+\nu_\ell$~\cite{Zhou:2019vxt}. Thus, the resonant component of these inclusive trident channels can be viewed as WBP followed by leptonic $W$ decay rather than as an entirely separate process.

To produce an on-shell $W$ boson, the neutrino-photon subsystem must satisfy the kinematic condition
$\hat{s} \geq (m_W+m_\ell)^2$. In the coherent and diffractive regimes, the effective threshold is controlled by the photon virtuality allowed by the corresponding nuclear and nucleon form factors. The nuclear form factor strongly suppresses coherent scattering for momentum transfers above $Q\sim 0.1~\mathrm{GeV}$, while diffractive scattering on individual nucleons can probe larger momentum transfers, up to $Q\sim 1~\mathrm{GeV}$~\cite{Zhou:2019vxt}. Consequently, in the $E_\nu\simeq 3$--$5~\mathrm{TeV}$ range relevant for MuCol$\nu$, the coherent contribution is strongly suppressed and the diffractive contribution remains subleading.

WBP at muon-collider neutrino energies is dominated by inelastic scattering. In this inelastic regime, the neutrino-photon invariant mass is approximately $\hat{s}\simeq 2xm_NE_\nu$, where $x$ is the fraction of the nucleon's momentum carried by the photon. This implies
$x_{\rm min}\simeq (m_W+m_\ell)^2/(2m_NE_\nu)\simeq m_W^2/(2m_NE_\nu)$.
In the limiting case $x\to1$, the minimum incident neutrino energy is therefore $E_\nu\gtrsim m_W^2/(2m_N)\simeq3.44~\mathrm{TeV}$. 
Therefore, a proposed future muon collider operating at a $10~\textrm{TeV}$ center-of-mass energy, capable of producing a neutrino flux with energies up to $5~\textrm{TeV}$, would be well suited to measure this yet unobserved process for the first time.
Over the energy range accessible at a $10~\mathrm{TeV}$ muon collider, $x_{\rm min}\simeq0.69$ at $E_\nu=5~\mathrm{TeV}$ and $x_{\rm min}\simeq0.86$ at $E_\nu=4~\mathrm{TeV}$, approaching unity near threshold. 
As a result, WBP with muon collider neutrinos probes the extreme large-$x$ partonic photon content of individual nucleons. In contrast, the additional quark-initiated contribution, after avoiding its overlap with the photon-initiated description, is negligible below $E_\nu\sim10^8~\mathrm{GeV}$~\cite{Zhou:2019vxt}.

As discussed in \cref{sec:nuatmuc}, although a pure silicon vertex detector is the baseline MuCol$\nu$ design, a realistic forward neutrino detector configuration may alternate active silicon tracking layers with passive layers of dense, heavy material to increase target mass and overall interaction rate. For illustration, we present in \cref{tab:W_events} the expected event rates in MuCol$\nu$ for pure $100~\text{kg}$ silicon and tungsten targets, assuming a 10-year exposure. As can be seen, the WBP rate is roughly an order of magnitude smaller than the dimuon trident signal in silicon discussed in \cref{sec:numutridents}. Notably, the expected event rate in silicon is slightly higher than in tungsten because silicon has a larger proton-to-neutron ratio. Therefore, comparing WBP rates between interleaved targets provides a novel method to disentangle the proton and neutron contributions. This would bypass the reliance on deuterium-based data~\cite{Xie:2023qbn} and improve modeling of photon-initiated cross sections on heavy, neutron-rich nuclear targets in high-energy colliders, beam-dump experiments, and neutrino telescopes~\cite{Zhou:2019frk,Xie:2023qbn}. 

\begin{table*}[tbp]
\centering
\caption{Expected number of WBP events ($\nu_\ell N \to \ell^- W^+ X$) for benchmark $100~\text{kg}$ Silicon (Si) and Tungsten (W) targets over a 10-year exposure at a $10~\text{TeV}$ muon collider. The total expected events are provided together with the relevant theoretical uncertainty, which is dominated by large-$x$ photon PDF unknowns. Event rates subdivided by the primary $W$ boson decay modes are also shown.}
\label{tab:W_events}
\resizebox{0.65\textwidth}{!}{%
\begin{tabular}{c|c||c||c|c}
\hline
\hline
Target & Mass & Total $W$ & Hadronic Decays & Leptonic Decays \\ 
nucleus & [kg] & Events & ($W \to q\bar{q}^\prime$) & ($W \to \ell \nu$) \\ 
\hline
Si & 100 & $(6.44\pm 5.96)\times 10^4$ & $4.34\times 10^4$ & $2.10\times 10^4$ \\ 
\hline
W & 100 & $(6.07\pm 5.60)\times 10^4$ & $4.09\times 10^4$ & $1.98\times 10^4$\\ 
\hline
\hline
\end{tabular}
}
\end{table*}

Experimentally, detecting WBP and disentangling it from both the overwhelming CC DIS background and pure neutrino trident production relies on the specific kinematics of the interaction. Because the momentum transfer remains constrained to $Q^2 \lesssim 10~\text{GeV}^2$ even for inelastic production~\cite{Zhou:2019vxt}, the energy transfer to the hadronic target is relatively soft, compared with the multi-TeV incident neutrino energy. 
The resulting soft hadronic activity provides a handle for signal identification, though the background rejection strategy depends on the decay mode of the $W$ boson:

\begin{itemize}
\item \textbf{Hadronic Decays ($W \to q\bar{q}^\prime$):} In this case, the final state consists of a charged lepton and energetic hadrons. The absence of missing energy enables precise reconstruction of the incident neutrino energy and allows for the isolation of the highest energy bins, $E_\nu \gtrsim 3.5~\text{TeV}$, where the on-shell $W$ production threshold is surpassed. Despite this restricted kinematic regime, the CC DIS background exceeds the signal by about $10^4$. Therefore, effective background rejection must rely on the specific kinematics of threshold $W$ production.

First, near the threshold, the massive $W$ boson absorbs almost the entire incident neutrino energy~\cite{Zhou:2019vxt}. Consequently, the primary charged lepton is relatively soft. In standard CC DIS, the outgoing lepton typically retains a large fraction of $E_\nu$, and an upper bound on the primary lepton energy may serve as the primary signal discriminant.

Second, the hadronic topologies differ significantly between signal and CC DIS background. The latter typically produce a high-multiplicity shower of energetic forward tracks. In contrast, $W$ production concentrates hadronic activity into a single highly boosted system. Because of the extreme boost, the $W$ boson decay products merge into a single ``fat jet'' characterized by a two-prong substructure and an invariant mass peaking at $m_W \approx 80.4~\text{GeV}$. The resonance can be identified by employing advanced jet-substructure techniques and $W$-tagging algorithms developed at the LHC, which currently achieve background rejection factors of $10^2$ to $10^3$ against light-quark and gluon jets~\cite{Larkoski:2017jix, CMS:2020poo}. Vetoing events with additional energetic tracks outside the primary jet cone will further help extract this rare signal from CC DIS backgrounds.

\item \textbf{Leptonic Decays ($W \to \ell \nu_\ell$):} Leptonic channels are more challenging for discovering WBP. 
The neutrino from $W$ decay carries away a significant fraction of the incident neutrino energy, which degrades its reconstruction and the identification of the $W$ boson resonance. 
Instead, this decay mode contributes to the neutrino trident rate. Specifically, in the dimuon channel, an incident $\nu_\mu$ produces a primary $\mu^-$, and the subsequent $W^+ \to \mu^+ \nu_\mu$ decay results in a visible $\mu^-\mu^+$ pair that closely resembles the dimuon trident signal.

Distinguishing this on-shell WBP from coherent (non-resonant) neutrino tridents should rely on hadronic recoil activity and dimuon kinematics. First, soft hadronic tracks serve as a signature of the inelastic nature of $W$ production, which is missing in coherent trident events. Second, the outgoing dimuon kinematics differ. Standard coherent tridents are characterized by extremely collinear muons~\cite{Altmannshofer:2024hqd}, while the secondary muon from $W$ decay acquires a large transverse momentum, $p_T \sim m_W/2$. This results in a much larger opening angle between the two muons, supporting event discrimination.

The flavor composition of the forward neutrino beam at the muon collider provides a further advantage for tau-lepton trident searches. Because the beam is dominated by $\nu_\mu$ and $\bar{\nu}_e$ and effectively lacks a primary $\nu_\tau$ component, charged-current scattering cannot produce a primary $\tau$ lepton. Consequently, on-shell WBP does not yield a $\tau^-\tau^+$ final state, and the corresponding non-resonant trident signal is free from the resonant WBP contribution.
\end{itemize}

Observing WBP near its kinematic threshold provides a novel opportunity to indirectly constrain fundamental hadronic physics. While the theoretical uncertainty on the cross section above $10~\text{TeV}$ is governed by the well-understood global photon parton distribution function (PDF)~\cite{Xie:2021equ} and is typically below $10\%$, the uncertainty in the $3$--$5~\text{TeV}$ threshold region can reach beyond $100\%$~\cite{Xie:2023qbn}. This is driven by the large-$x$ photon PDF of the nucleon, especially the neutron, whose structure functions rely on highly uncertain nuclear corrections required to extract free-nucleon data from deuterium scattering experiments~\cite{Bosted:2007xd,Ye:2017gyb,Accardi:2016qay}. Furthermore, the cross section relies heavily on the theoretical modeling of low-$Q^2$ hadronic interactions. The latter includes empirical parameterizations of resonance structure functions~\cite{CLAS:2003iiq,Christy:2007ve}, purely kinematic target-mass corrections (TMCs)~\cite{Georgi:1976ve,Schienbein:2007gr}, and dynamical higher-twist effects reflecting multi-parton correlations~\cite{Abt:2016vjh,Accardi:2016qay, Amoroso:2022eow}. A precision measurement of WBP would serve as a new indirect tool to reduce these uncertainties.

\begin{figure}[!t]
    \centering
    \includegraphics[width=\columnwidth]{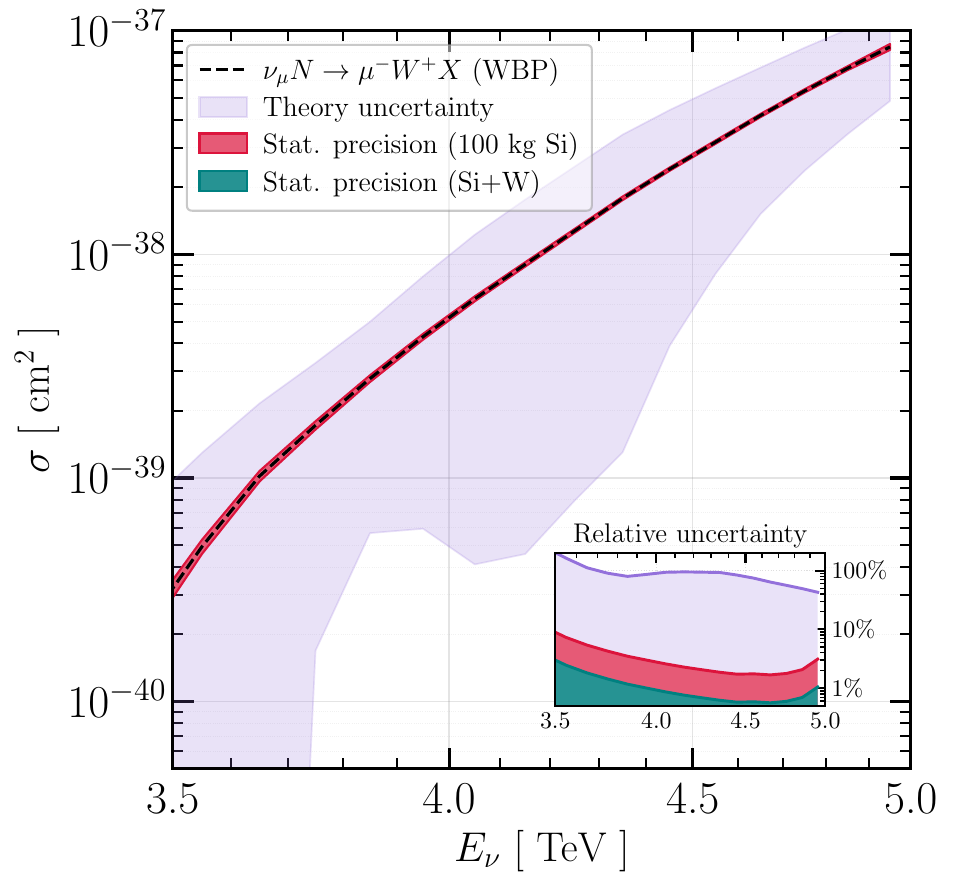}
    \caption{Projected sensitivity to WBP, $\nu_\mu N \to \mu^- W^+ X$, at a $10~\text{TeV}$ muon collider compared to present theoretical uncertainties. The main panel shows the cross section as a function of incident neutrino energies available at MuCol$\nu$. The shaded purple band indicates the current theoretical uncertainty, driven by the large-$x$ photon PDF and non-perturbative hadronic dynamics. The red overlay corresponds to the projected statistical uncertainty for a $100~\text{kg}$ active silicon target, assuming a 10-year exposure and evaluated in $100~\text{GeV}$ incident-energy bins. It is obtained based on the hadronic decay channel, $W \to q\bar{q}^\prime$. The inset illustrates the relative uncertainty, $\delta\sigma/\sigma$, for both the pure silicon target (red) and a combined detector configuration with interleaved tungsten layers (green), compared with the theoretical band (purple).}
   \label{fig:wbp}
\end{figure}

To illustrate this, we show in \cref{fig:wbp} the projected sensitivity to the WBP cross section against present theoretical uncertainties. The shaded purple band indicates the theoretical uncertainty, which grows beyond $100\%$ near the kinematic threshold. We overlay this with sensitivity projections for MuCol$\nu$, based on a 10-year exposure at a $10~\textrm{TeV}$ collider, which corresponds to an incident neutrino flux up to $5~\textrm{TeV}$. We focus exclusively on $W$ boson hadronic decays, $W \to q\bar{q}^\prime$, that allow for precise neutrino energy measurement, as discussed above. We consider neutrino-energy bins of $100~\textrm{GeV}$ width. The projected sensitivity for the $100~\textrm{kg}$ silicon detector is shown in red, indicating an expected order-of-magnitude gain in precision. A combined detector, incorporating an interleaved tungsten target with the same transverse size and length, corresponding to approximately $842~\text{kg}$ of tungsten, can achieve further improvement, up to sub-percent accuracy (green shading). Realistic projections should further include systematic detector effects, finite reconstruction efficiencies, and beam flux uncertainties. 
For analyses exploiting charged-lepton and shower-energy distributions to suppress CC DIS backgrounds, QED final-state radiation should also be taken into account, as it can modify these distributions~\cite{Plestid:2024bva}.
Hence, measuring WBP in MuCol$\nu$ will pave the way for constraining the relevant cross section far beyond current theoretical understanding.

Because the production threshold requires a minimum momentum fraction $x_{\text{min}} \approx m_W^2 / (2 m_N E_\nu)$, studying the event rate as a function of the reconstructed incident neutrino energy, $E_\nu \approx E_\ell + E_W$, effectively maps the $x$-dependence of the photon PDF. Ultimately, a precision measurement of the total inclusive WBP cross section will constrain the $Q^2$-integrated LUXqed photon flux~\cite{Manohar:2016nzj,Manohar:2017eqh}, severely restricting the available parameter space for modeling these non-perturbative hadronic effects, cf. further discussion in Refs~\cite{Xie:2021equ,Xie:2023qbn}. 

Constraining the large-$x$ photon PDF and low-$Q^2$ non-perturbative models has further implications beyond the muon collider. At future hadron colliders, e.g., the proposed FCC-hh, it would drastically shrink theoretical uncertainties for high-mass photon-initiated backgrounds, such as TeV-scale $W^+W^-$ and $WH$ production~\cite{Manohar:2016nzj,Xie:2021equ}. This will also allow for improved legacy re-interpretations of LHC data. Furthermore, the measurement will provide calibration data for the low-$Q^2$ hadronic models that dominate the systematic uncertainties of next-generation neutrino oscillation experiments operating in the few-GeV regime~\cite{NuSTEC:2017hzk}.

Finally, on the BSM side, this measurement will reduce theoretical uncertainties for axion-like particles (ALPs) produced via photon fusion. While forward LHC experiments like FASER primarily targeted light ALPs generated through coherent Primakoff scattering where partonic uncertainties are subdominant~\cite{Feng:2018pew,FASER:2024bbl}, a large-$x$ photon PDF is essential for searching for heavier ALPs at future colliders and beam-dump facilities~\cite{Cesarotti:2022ttv,Xie:2023qbn}. 

\section{Bounds on four-fermi muon interactions}
\label{sec:BSMbounds}

Beyond testing the electroweak sector to unprecedented precision, the massive event samples at the proposed MuCol$\nu$ detector would turn neutrino tridents into a sensitive probe for BSM physics. Because the dimuon trident cross section relies on the destructive interference between charged- and neutral-current diagrams, even minuscule modifications to the leptonic gauge couplings or the introduction of new heavy mediators can measurably alter the interaction rate. Following Ref.~\cite{Altmannshofer:2019zhy}, the weak interaction in trident events is described via the effective Hamiltonian,
\begin{align}
\mathcal{H}_{\rm eff} =
\frac{G_F}{\sqrt{2}} \bigg(
  &g_V^{ijkl} (\bar{\nu}_i \gamma_\alpha P_L \nu_j)(\bar{\ell}_k \gamma^\alpha \ell_l)\nonumber\\
+ &g_A^{ijkl} (\bar{\nu}_i \gamma_\alpha P_L \nu_j)(\bar{\ell}_k \gamma^\alpha \gamma_5 \ell_l) \bigg) ~,
\end{align}
since the momentum transfer is typically well below the electroweak scale. Here $G_F$ is the Fermi coupling, $P_L$ is the left-handed projection operator, $\gamma_\alpha$ are the Dirac gamma matrices, $g_{V/A}^{ijkl}$ are the (axial-) vector couplings, and $i,j,k,l$ index the lepton flavors. New physics interfering with muonic interactions is parameterized by modifying the vector and axial-vector couplings $g_V^{\mu\mu\mu\mu} \equiv g_V$ and $g_A^{\mu\mu\mu\mu} \equiv g_A$, such that $g_V = 1 + 4\sin^2\theta_\mathrm{W} + \Delta g_V$ and $g_A = -1 + \Delta g_A$, with $\theta_\mathrm{W}$ being the Weinberg angle.

To project the BSM sensitivity, we assume that prompt and beam-induced backgrounds can be suppressed to a negligible level using the kinematic and topological handles discussed in \cref{sec:numutridents}. Furthermore, we assume that the resonant WBP contribution can be successfully disentangled from the non-resonant trident events following the strategies outlined in \cref{sec:Wboson}. Hence, we focus exclusively on the coherent, non-resonant dimuon ($\mu\mu$) trident signal for the BSM sensitivity study, as this channel provides the cleanest experimental signature. 

However, because the detector cannot distinguish the flavor of the initial and final-state neutrinos on an event-by-event basis, the total expected SM event rate must be evaluated inclusively. We therefore compute the SM baseline by summing these non-resonant dimuon trident contributions from all incoming neutrino flavors present in the primary beam, which is dominated by $\nu_\mu$ and $\bar{\nu}_e$. Because our BSM parameterization specifically targets modifications to the $\nu_\mu$--$\mu$ coupling, $\Delta g_{V/A}^{\mu\mu\mu\mu}$, the $\bar{\nu}_e$-initiated tridents are assumed to remain at their SM rates, and they contribute as an irreducible baseline to the total observed event count.

\begin{figure}[!t]
    \centering
    \includegraphics[width=0.47\textwidth]{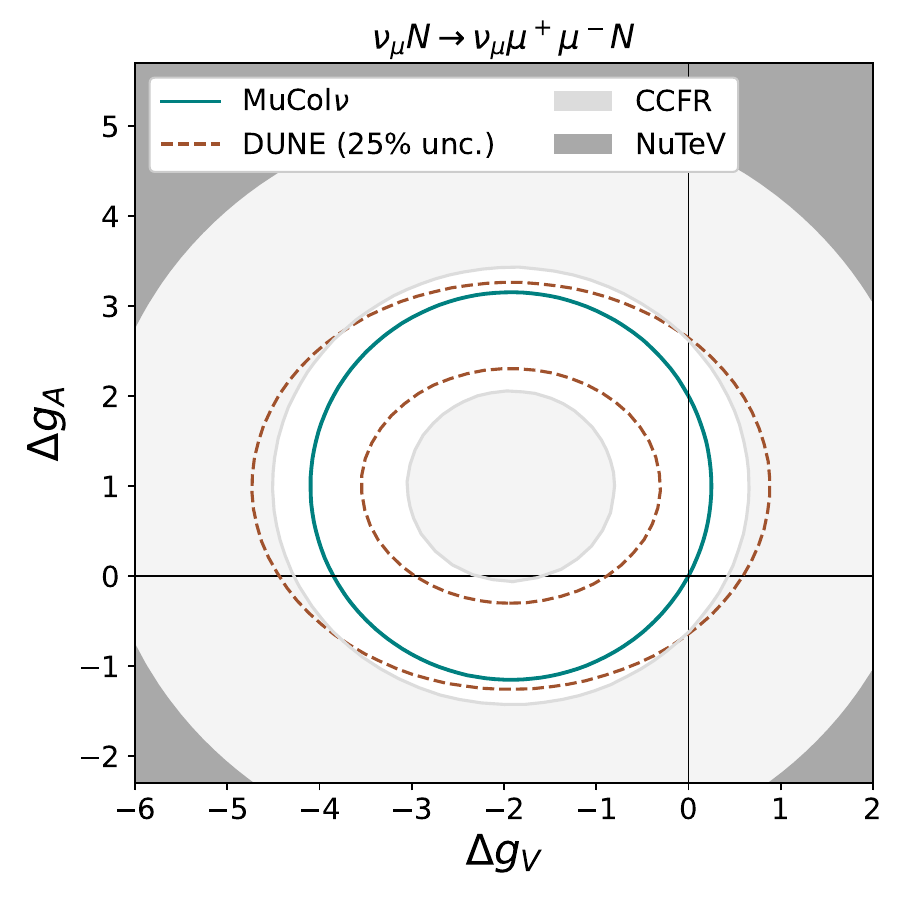}
    \caption{Constraints on modifications to the axial and vector couplings from CCFR and NuTeV measurements (gray), compared to the projected bounds from DUNE assuming a 25\% uncertainty cross section measurement (dashed brown). For the latter, the area between the two ellipses remains allowed. The NuTeV bound is computed following Ref.~\cite{Altmannshofer:2019zhy} and assumes equal BSM-to-SM cross-section ratios at CCFR and NuTeV due to their similar energies. The bounds obtained with the proposed silicon detector at the muon collider (teal) are extremely stringent and practically overlap, forming a single circle of allowed points that passes through the crossing of the solid black lines at $(\Delta g_V,\Delta g_A)=(0,0)$, corresponding to the SM. All sensitivities are at 95\% confidence level.}
    \label{fig:FFconstraint}
\end{figure}

Conservatively, we assume the total number of observed events follows a simple Poisson distribution integrated over the entire $E_\nu$ range of the experiment, without performing a differential shape analysis. The constraints shown in \cref{fig:FFconstraint} are event-rate contours in the $(\Delta g_V, \Delta g_A)$ plane, representing the 95\% confidence-level deviations above and below the expected SM event rate. In principle, this procedure yields upper and lower contours that define an allowed annular region for $\Delta g_V$ and $\Delta g_A$. However, because the expected trident statistics are so extraordinarily high at the MuCol$\nu$ detector, the resulting bounds become extremely stringent. The two contours effectively merge into a single ring that passes through the SM expectation at $(\Delta g_V,\Delta g_A)=(0,0)$. Hence, the neutrino trident measurement at MuCol$\nu$ will constrain virtually all parameter space of this scenario.

\section{Conclusions}
\label{sec:conclusions}

In addition to serving as precision machines for new physics searches at the energy frontier, proposed muon colliders are exceptionally bright sources of high-energy neutrinos. Muon decays within the straight sections of the storage ring produce intense, highly collimated neutrino beams with multi-TeV energies and a well-understood flavor composition. With an expected yield of approximately $2 \times 10^{10}$ deep inelastic scattering events over ten years in a compact $100~\text{kg}$ target located $500~\text{m}$ downstream, this facility offers an unprecedented environment where even highly suppressed, rare electroweak processes are expected to occur at high rates.

In this study, we investigated the feasibility of measuring both non-resonant neutrino trident production and on-shell $W$-boson production for the first time. We project that the baseline active silicon target of the proposed MuCol$\nu$ detector will yield nearly $10^6$ inclusive trident events and more than $10^4$ WBP events. By isolating purely leptonic trident signatures, we showed that muon tridents will enable rigorous Standard Model tests, setting limits on new vector and axial-vector couplings in weak effective field theory that significantly surpass the projected sensitivity of DUNE. Concurrently, isolating the WBP signal near its kinematic threshold will indirectly constrain the nucleon's large-$x$ photon parton distribution function with unprecedented precision, drastically reducing theoretical uncertainties for future hadron collider backgrounds and low-energy neutrino oscillation models. 

Despite these large expected event yields, trident and WBP signals remain exceptionally rare, accounting for roughly one in $10^5$ neutrino interactions. Measuring them depends critically on effective background rejection, particularly against inclusive deep inelastic scattering, single-pion production, and diffractive charm production. Consequently, we have identified several key detector requirements to fully exploit this physics potential: robust lepton-hadron separation; high-precision vertexing with micrometer resolution to identify and reject short-lived particle decays, such as $D_s^\pm$ mesons and $\tau$ leptons; excellent missing transverse momentum resolution; and sufficient calorimetric containment to reliably veto hadronic activity. These requirements should guide detector design to fully exploit the physics potential of neutrino experiments at a muon collider.

Furthermore, alternating active silicon tracking layers with passive high-$Z$ target materials, such as tungsten, would not only increase the overall interaction rate but also provide a novel method to disentangle proton and neutron contributions to the WBP cross section. Together, these possible measurements would significantly strengthen the already extensive physics case for a dedicated forward neutrino experiment at a future muon collider.

\section*{Acknowledgements}

The work of T.\,M. is supported in part by U.S. National Science Foundation Grants PHY-2111427 and PHY-2210283 and Heising-Simons Foundation Grant 2020-1840. S.\,T. was supported by the National Science Centre, Poland (research grant No. 2021/42/E/ST2/00031). S.\,T. was also partially supported by the FNP IRA programmes: AstroCeNT (MAB/2018/7), funded from the ERDF, and Astrocent (FENG.02.01-IP.05-A015/25) co-financed by the European Union under FENG 2021–2027; and Teaming for Excellence grant Astrocent Plus (101137080) funded by the European Union with complementary national funding from the MNiSW (MNiSW/2025/DIR/811). B.\,Z. was supported by Fermi Forward Discovery Group, LLC under Contract No. 89243024CSC000002 with the U.S. Department of Energy, Office of Science, Office of High Energy Physics.


\clearpage
\bibliography{references}

\end{document}